# Quantum critical behavior induced by Mn impurity in $CuGeO_3$

A.V. Semeno[a] [*], N.E. Sluchanko[a], N.A. Samarin[a], A.A. Pronin[a], H. Ohta[b], S. Okubo[b] and S.V. Demishev

[a]*Low Temperatures and Cryogenic Engineering Department, A.M. Prokhorov General Physics Institute of Russian Academy of Sciences, Vavilov street, 38, 119991 Moscow, Russia*

[b]*Molecular Photoscience Research Center, Kobe University, 1-1 Rokkodai, Nada, Kobe 657-8501, Japan*

**Abstract**

Results of high frequency (60-315 GHz) studies of ESR in $CuGeO_3$ single crystals containing 0.9% of Mn impurity are reported. Quantitative EPR line shape analysis allowed concluding that low temperature magnetic susceptibility for $T<40$ K diverges as $\chi \sim 1/T^{\alpha}$ with the critical exponent $\alpha=0.81\pm0.03$ and therefore manifests onset of a quantum critical (QC) regime. We argue that transition into Griffiths phase occurs at $T_G \sim 40$ K and disorder produced by Mn impurity in quantum spin chains of $CuGeO_3$ may lead to co-existence of the QC regime and spin-Peierls dimerisation.

Recently it has been shown that doping of the spin-Peierls compound $CuGeO_3$ with magnetic impurities such as iron and cobalt gives rise to a disorder driven quantum critical (QC) phenomena [1-3]. The insertion into antiferromagnetic $Cu^{2+}$ quantum spin chains (S=1/2) of a magnetic ions with S=3/2 (Co) or S=2 (Fe) on a concentration level $x$=1-2% have lead to the damping of both spin-Peierls and Neel transitions at least down to 1.8 K or 0.5 K in the cases of Co and Fe respectively [4,5]. In the interval $T<T_G \sim 40$ K the magnetic susceptibility of $CuGeO_3$:M (M=Fe, Co) have acquired power law

$$\chi(T) \sim 1/T^{\alpha}, \qquad (1)$$

with the exponent $\alpha<1$, which is characteristic to a Griffiths phase [6,7].

From the other hand, Ni impurity (S=1) in $CuGeO_3$ is known to follow a standard scenario of doping [8] and in aforementioned concentration range spin-Peierls state coexisting with the Neeel state were observed [9]. In this situation it is interesting to investigate behavior of Mn (S=5/2) impurity. To our best knowledge the only paper where $CuGeO_3$:Mn system have been mentioned so far is Ref. 10 but no details about physical properties of this material have been reported.

In the present paper we report results of high frequency 60-315 GHz electron spin resonance (ESR) study of $CuGeO_3$ single crystals containing 0.9% of Mn impurity in magnetic field $B$ up to 12 T at temperatures 1.8-70 K. The quality of samples has been controlled by X-ray and Raman scattering data; the actual content of impurity in the sample has been checked by means of chemical analysis. The details about synthesis technique can be found elsewhere [11]. In ESR experiments magnetic field was alined along ***a*** crystallographic axis.

The example of the observed ESR spectra is presented in fig. 1. It is visible that there is a single lorentzian line, which broadens and shifts in a non-monotonous way when temperature is lowered. The quality of spectra have allowed to perform line shape analysis and obtain all spectroscopic characteristics (line width, *g*-factor and integrated intensity) as a function of temperature. It is worth to note that for $T>70$ K the *g*-factor value was $g \approx 2.1$, which is close to $g=2.15$ characteristic to the ***B***||***a*** in undoped $CuGeO_3$ [1-5]. Therefore it is possible concluding that absorption line in fig. 1 is caused by ESR on $Cu^{2+}$ chains modified by Mn impurity. Below we will consider temperature dependence of integrated intensity $I(T) \sim \chi(T)$ and the rest parameters will be discussed in a separate publication.

---

[*] Corresponding author. Tel.: +7-095-132-8253; fax: +7-095-135-8129; e-mail: semeno@lt.gpi.ru.

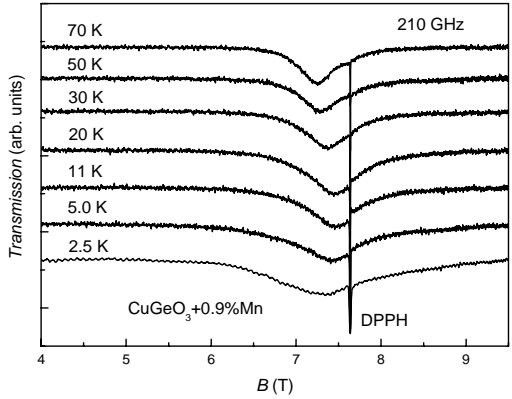

Fig. 1. ESR spectra obtained for 210 GHz using quasi-optical technique.

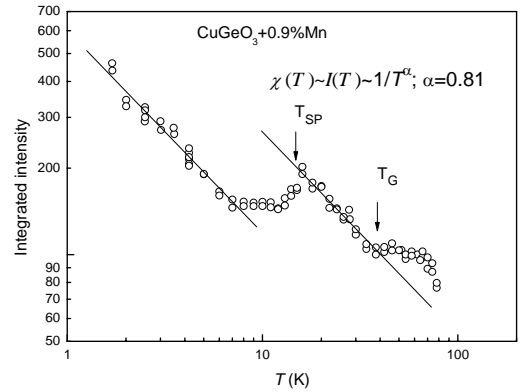

Fig. 2. Temperature dependence of the integrated intensity for 210 GHz. Solid lines represent power asymptotic of magnetic susceptibility (Eq. (1)). Arrows mark transitions into Griffiths ($T_G$) and spin-Peierls state ($T_{SP}$).

It follows from fig. 2 that power asymptotic of magnetic susceptibility with the exponent $\alpha=0.81\pm0.03$ specific to disorder driven QC regime begins at temperature $T_G \sim 40$ K, which is in agreement with the observed in $CuGeO_3$:Fe and $CuGeO_3$:Co temperatures for the transition into Griffiths phase [1-5]. However at spin-Peierls transition temperature $T_{SP}=14.5$ K a sharp decrease of integrated intensity occur manifesting partial dimerisation of $Cu^{2+}$ spin chains. For T<7 K the power law with the *same* $\alpha$ is restored and holds up to the lowest temperature studied (fig. 2). Interesting that observed value $T_{SP}=14.5$ K is equal to that in pure $CuGeO_3$ crystal [11].

A condition $T_{SP}<T_G$ means that spin-Peierls transition in $CuGeO_3$:Mn occur within Griffiths phase, i.e. when magnetic subsystem is divided into spin clusters with different coupling constants [6-7]. It is reasonable to suppose that lowering temperature will lead to increase of the cluster size [6-7] and a transition into the dimerised state may be allowed for the chains inside the cluster if its size at $T=T_{SP}$ will big enough. In this scenario it is possible expecting dispersion of $T_{SP}$ for various spin clusters; however the distribution of $T_{SP}$ will be restricted from above by spin-Peierls transition temperature in pure sample. Therefore transition into dimerised state will begin at $T_{SP}$ in undoped case as observed experimentally (fig. 2). Below $T_{SP}$ the magnetic contribution from dimerised chains will vanish rapidly and power law for susceptibility may be restored. As long as in disorder driven QC regime index $\alpha$ depends on only space dimension and dynamic exponent, which connects time and length scales [12], it is natural to expect restoration of the same $\alpha$ value at $T<T_{SP}$ in agreement with the experimental data (fig. 2).

In conclusion we have shown that disorder produced by Mn impurity in quantum spin chains of $CuGeO_3$ may lead to co-existence of the QC regime and spin-Peierls dimerisation. The possibility of this scenario have not been foreseen up to now and more theoretical studies are required to clarify behavior of different magnetic impurities in a spin-Peierls system.

Support from RFBR grant 04-02-16574 and programme of RAS "Strongly correlated electrons" is acknowledged. This work was partly supported by a Grant-in-Aid for Scientific Research on Priority Areas (No. 13130204) from the ministry of Education, Culture, Sports, Science and Technology of Japan.